\documentclass [10pt, aps, prd, amsmath, floats, floatfix, twocolumn, notitlepage,
superscriptaddress, nofootinbib, showpacs, longbibliography]{revtex4-2}
\usepackage[T1]{fontenc}
\usepackage[utf8]{inputenc}
\usepackage{lmodern}
\usepackage{verbatim}
\usepackage{physics}
\usepackage{orcidlink}
\definecolor{linkcolor}{rgb}{0.6, 0.0, 0.0}
\usepackage{cleveref}  
\usepackage[all]{hypcap}
\usepackage{graphicx}
\usepackage{xspace}
\usepackage{amssymb}
\usepackage{amsmath}
\usepackage[normalem]{ulem} 
\usepackage{bm} 
\usepackage{microtype}
\usepackage[english]{babel}
\usepackage{blindtext}
\usepackage{array}
\usepackage{tabularx}
\usepackage{multirow}
\usepackage{subfigure}

\usepackage{natbib}
\graphicspath{%
  {figs/}%
}

\DeclareMathOperator\arctanh{arctanh}

\newcommand{\scr}{\scriptscriptstyle}

\begin{document}

\title{Gravitational collapse and singularity avoidance of a
homogeneous dust fluid on a brane with timelike extra dimension}


\author{Rikpratik Sengupta}
\email{rikpratik.sengupta@gmail.com}
\affiliation{Department of Physics, Indian Institute of Technology Kanpur, Kalyanpur 208016, India}

\author{Chiranjeeb Singha}
\email{chiranjeeb.singha@iucaa.in}
\affiliation{Inter-University Centre for Astronomy \& Astrophysics, Post Bag 4, Pune 411 007, India}

\begin{abstract} 
We investigate the gravitational collapse of a homogeneous dust fluid in the Shtanov-Sahni model consisting of a braneworld possessing an extra time-like dimension. A Friedmann-Lemaître spacetime models the interior of the collapsing cloud, while the exterior consists of a Vaidya radiation envelope, which eventually transitions to a static vacuum described by the Reissner–Nordström (RN) solution with a positive tidal charge. Although a smooth matching between the interior and the static vacuum is not possible due to the breakdown of Birkhoff’s theorem in the braneworld context, we demonstrate that as long as braneworld effects remain significant, the brane tension remains finite, ensuring the scalar curvature to be bounded, thus avoiding the formation of a singularity.
\end{abstract}

\maketitle
    
\section{Introduction}	

Gravitational collapse is a fundamental process in astrophysics that determines the final fate of massive stars once their nuclear fuel is exhausted. When the internal pressure is no longer sufficient to counterbalance gravitational attraction, the star undergoes a dramatic contraction. Depending on the star’s mass, this collapse can lead to the formation of compact remnants such as white dwarfs, neutron stars, or black holes. Within the framework of general relativity, gravitational collapse is not merely a dynamical process; it signifies a deep geometrical restructuring of spacetime, potentially giving rise to horizons and singularities where classical descriptions of physics break down.

Oppenheimer and Snyder carried out the first exact relativistic analysis of such a collapse in 1939 \cite{oppenheimer1939}. Their model, now known as the Oppenheimer-Snyder (OS) solution, considers the idealized scenario of a spherically symmetric, homogeneous dust cloud undergoing collapse. Assuming vanishing pressure and perfect spherical symmetry, they employed a closed Friedmann–Lemaître–Robertson-Walker (FLRW) metric to describe the cloud’s interior evolution under its own gravity. The exterior region was matched to the Schwarzschild vacuum solution at the boundary of the collapsing matter. Their work showed that such a collapse inevitably leads to the formation of an event horizon and a central singularity, thereby providing a compelling description of black hole formation. Despite its idealized nature, the OS model remains a cornerstone in gravitational physics due to its analytical simplicity and its clear demonstration of relativistic collapse.

Adopting an FLRW interior is well-motivated: the assumptions of homogeneity and isotropy naturally lead to an FLRW geometry as a solution to Einstein’s field equations with pressureless matter. Moreover, the spherical symmetry and lack of pressure ensure that Birkhoff’s theorem applies, justifying the use of the Schwarzschild metric in the exterior. The closed (positively curved) FLRW model is particularly suitable, as it represents a spatially finite spherical region, consistent with the physical configuration of the collapsing dust cloud.

Nevertheless, the OS model relies on strong idealizations, particularly the assumption of a uniform matter distribution, which is unlikely to be realized in realistic astrophysical conditions. A more general family of solutions, known as the Lemaître–Tolman–Bondi (LTB) models, relaxes the condition of homogeneity while preserving spherical symmetry and zero pressure \cite{tolman1934,bondi1947}. In these models, the matter density is allowed to vary with the radial coordinates, leading to an inhomogeneous collapse in which different shells contract at different rates. This inhomogeneity introduces a richer causal structure and allows for more intricate singularity formation, including the possibility of naked singularities under certain initial configurations. As a result, LTB models have become essential tools for exploring the cosmic censorship conjecture and the nonlinear behavior of gravitational collapse.

The OS and LTB models provide profound insight into the end-states of massive stars and the geometry of a dynamical spacetime. Though based on simplifying assumptions, they successfully capture key qualitative features of general relativistic collapse and continue to serve as critical benchmarks for more sophisticated scenarios involving pressures, angular momentum, or quantum effects.

Several recent works have explored extensions of the classical Oppenheimer-Snyder collapse to incorporate quantum gravitational effects and model nonsingular, bouncing compact objects. In \cite{GCB1}, a thin-shell approach is used to generalize the collapse scenario without committing to a specific mechanism for singularity resolution. The analysis demonstrates that a bounce must occur at or below the horizon formation threshold, thus avoiding the formation of trapped regions under minimal assumptions. This implies that consistent black-to-white hole bounce models may necessitate additional structure, such as an inner horizon. Building on this idea, \cite{GCB2} employs effective dynamics from loop quantum cosmology (LQC) to model the bounce, matching it to an exterior Schwarzschild geometry with minimal junction conditions. The resulting model features no trapped region and describes a pulsating compact object with both UV and IR cutoffs. While it does not model black-to-white hole transitions, it offers a consistent framework for Planck-scale relics, excluding macroscopic stellar bounces. A more general, model-independent perspective is provided in \cite{GCB3}, which derives kinematical constraints showing that a bounce must occur in an untrapped region or along a trapping horizon. The work identifies additional dynamical conditions required for effective bounce models and introduces a family of solutions characterized by both classical and quantum parameters. These include bouncing stars, bouncing black holes, and hybrid objects, with an explicit black-to-white hole bounce constructed using spatially closed LQC. 

Despite differences in their formulations, the three models in \cite{GCB1, GCB2, GCB3} converge on several key insights. At their core, each replaces the classical singularity of gravitational collapse with a bounce, allowing for a non-singular evolution of the collapsing matter. A notable consequence shared across these models is the absence of a permanently trapped region; if such a region forms at all, it is only transient or marginal, occurring at most on a trapping horizon. This precludes the development of a classical black hole interior and suggests alternative end states for collapse. The models employ minimal matching conditions, typically connecting a modified (quantum or effective) interior to a classical Schwarzschild exterior via the continuity of the metric across a timelike surface. While a simple bounce scenario is broadly compatible with these assumptions, the realization of a full black-to-white hole transition requires additional structural elements, such as an inner horizon or more detailed quantum corrections, to maintain consistency. Importantly, the physical behavior and interpretation of the bounce depend on several key parameters. These include standard classical quantities, such as mass and radius, as well as quantum parameters introduced through ultraviolet completions, which determine whether the outcome resembles a pulsating object, a bounce near or beyond the Schwarzschild radius, or a transition between black hole and white hole phases. Some constraints, particularly those governing the location of the bounce relative to the formation of trapped regions, emerge as model-independent results, arising directly from geometric and junction conditions. Collectively, these models offer versatile frameworks that not only deepen our theoretical understanding of non-singular collapse but also provide a foundation for future investigations into the phenomenology of quantum gravity in astrophysical contexts. These shared conclusions point toward a coherent picture where quantum gravity-inspired corrections to gravitational collapse lead to non-singular, horizonless (or horizon-marginal) outcomes with potentially observable phenomenology.

Braneworld scenarios have attracted significant attention in recent years. A particularly influential model was introduced by Randall and Sundrum (RS) in 1999 \cite{Randall1,Randall2}, offering a potential solution to the hierarchy problem in particle physics \cite{VSK}. This framework addresses the relative weakness of gravity compared to the electroweak interaction by proposing that our universe is a $(3+1)$-dimensional brane embedded within a higher-dimensional bulk spacetime. While the fields of the Standard Model are confined to the brane, gravity is allowed to propagate into the bulk, which possesses the geometry of a five-dimensional Anti-de Sitter space ($AdS_5$). A negative cosmological constant in the bulk ensures the localization of gravity near the brane. The original RS model involved two branes, but in a subsequent paper \cite{Randall2}, the negative-tension brane was pushed to infinity, resulting in the RS-II model with a single brane.

This single-brane configuration, featuring a spacelike extra dimension, has found wide applications in both cosmological \cite{Binetruy,Maeda,Langlois,Chen,Kiritsis,Campos,Sengupta1,Maartens,SG1,SG2} and astrophysical~\cite{Wiseman2,Germani,Deruelle,Wiseman,Visser,Creek,Pal,Bruni,Govender,Sengupta5} contexts. A corresponding dual model was later proposed by Shtanov and Sahni (SS), in which the extra dimension is timelike rather than spacelike \cite{Sahni4}. This alters the bulk signature from Lorentzian to one with signature $(-,-,+,+,+)$. In contrast to RS-II, the SS model features a brane with negative tension embedded in bulk with a positive cosmological constant.

A particularly remarkable feature of the SS model is the natural emergence of a non-singular bounce during the contracting phase of the universe, even when the brane matter satisfies the usual energy conditions. There are no fundamental obstructions preventing such bounces from recurring indefinitely. When a massive scalar field is present on the brane, each bounce can increase in amplitude, offering a mechanism for addressing the flatness problem \cite{S6,S7,S8}. A closed universe filled with matter satisfying the strong energy condition (SEC) can thereby undergo a cyclic evolutionary trajectory. More generally, if the brane matter satisfies the null energy condition (NEC), non-singular bounces can occur in both the past and future. One may refer to \cite{Pat1, Pat2} for comprehensive reviews on different bouncing cosmological models and the physical mechanisms underlying the resolution of the singularity, marked by a smooth transition from the contraction phase to the currently observed expanding phase.

One complication in such scenarios is the tachyonic behavior of Kaluza–Klein gravitons in timelike extra-dimensional frameworks, as noted in \cite{Sahni4}. This issue was addressed in \cite{S10}, where a solitonic 3-brane solution was presented in the context of the five-dimensional Einstein–Hilbert–Gauss-Bonnet theory with spacetime signature $(-,+,+,+,-)$, featuring a second timelike direction. The brane, treated as a $\delta$-function source, completely localizes gravity, allowing no bulk propagating modes. Gravity on the brane is thus purely four-dimensional and governed by Einstein's theory of general relativity. Importantly, despite the presence of a timelike extra dimension, no tachyonic or negative-norm states arise. Furthermore, the feasibility of constructing stable, traversable wormholes within the SS braneworld framework has been examined \cite{RSM}. It was found that ordinary matter satisfying all classical energy conditions can source such wormholes on the brane, and that the Weyl curvature singularity at the wormhole throat, associated with infinite tidal forces, can be resolved by extra-dimensional effects.

Within the Randall–Sundrum braneworld framework, an exact black hole solution can be constructed on a three-brane \cite{ND1}, where the Reissner–Nordström metric arises without electric charge. Instead, a negative tidal charge stemming from extra-dimensional effects modifies the Schwarzschild potential, resulting in a single event horizon located outside the classical Schwarzschild radius. This solution, derived from a closed system of brane-localized equations, captures the strong gravity regime and includes features not present in General Relativity (GR). 

In scenarios involving the gravitational collapse of homogeneous dust on the brane, modeled after the Oppenheimer–Snyder framework, a no-go theorem indicates that the exterior spacetime cannot remain static, unlike in standard GR \cite{Bruni}. Further studies show that such a collapse must be accompanied by a Vaidya radiation envelope, mediating between the interior solution and the Reissner–Nordström exterior. This radiative layer is a distinct consequence of braneworld effects, implying that collapse on the brane necessarily involves null radiation, a behavior fundamentally different from classical GR predictions \cite{Govender}.

We adopt a similar approach to study gravitational collapse within the SS brane scenario. Given that the FRW cosmological solution on the SS brane avoids an initial singularity, it becomes a compelling question whether the collapse of a homogeneous dust sphere in this framework leads to the formation of a singularity in the future or not. 

This paper is organized as follows. In Section \ref{basic}, we outline the basic framework for the brane-world scenario. Section \ref{collapse} is devoted to analyzing the gravitational collapse of a homogeneous dust fluid in the Schwarzschild-de Sitter (SS) brane context. In Section \ref{collapsehalts}, we examine the post-collapse dynamics and explore the conditions under which the collapse may halt. Finally, Section \ref{conclusion} summarizes our findings and discusses possible directions for future research.

\noindent \textbf{Notations and Conventions:} \\
Throughout this paper, we use the metric signature $(-,+,+,+)$ for the $3+1$ spacetime and $(-,+,+,+)$ for the embedding 5-dimensional bulk. We also adopt natural units by setting the Newton constant and the velocity of light to unity, i.e., $G_\textsc{n} = c = 1$.

\section{Basic framework}\label{basic}

We consider a 4-dimensional brane $\mathcal{B}$ embedded in a
5-dimensional bulk $\mathcal{M}$ with two timelike coordinates. Denoting
the extra timelike coordinate by $\tau$ and setting the brane
$\mathcal{B}$ at $\tau =0$, the line element is assumed to take the
form (capital Latin indices span $\mathcal{M}$ while Greek ones are 
restricted to $\mathcal{B}$)
\begin{align}
g_{\scr AB}
\dd x^{\scr A } \dd x^{\scr B} = -\dd\tau^2 + 
e^{2\sigma\tau/(3M^3)}
h_{\mu\nu}\qty(x^\alpha) \dd x^\mu \dd x^\nu,
\label{ds5}
\end{align}
with $\sigma <0$, solving the Shtanov–Sahni (SS) action, which
reads, in terms of the 5 and 4-dimensional scalar curvatures
$\mathcal{R}$ and $R$,
\begin{align}
\begin{split}
\mathcal{S} = &M^3 \qty[ \int_\mathcal{M} \sqrt{g} \dd^5x 
\qty(\mathcal{R} -2 \Lambda_5) +2\int \sqrt{-h} \dd^4x K]\\ 
& + \int \sqrt{-h} \dd^4x \qty( m^2 R -2\sigma + \mathcal{L}),
\end{split}
\label{act5act4}
\end{align}
where $\Lambda_5$ is the bulk cosmological constant, $\sigma$
the 4-dimensional brane tension while $M$ and $m$ respectively
denote the 5-dimensional and 4-dimensional Planck masses. In
Eq.~\eqref{act5act4}, we suppose no bulk matter content, so
that variations with respect to $g_{\scr AB}$ yields
vacuum Einstein field equations (EFE)
\begin{align}
\mathcal{G}_{\scr AB} + \Lambda_5 
g_{\scr AB}=0
\label{EFE5}
\end{align}
for the 5-dimensional Einstein tensor 
$\mathcal{G}_{\scr AB} =
\mathcal{R}_{\scr AB} -\frac12 
g_{\scr AB} \mathcal{R}$ obtained from
the Ricci tensor 
$\mathcal{R}_{\scr AB}
= g^{\scr CD}
\mathcal{R}_{\scr CADB}$
and scalar $\mathcal{R} = g^{\scr AB}
\mathcal{R}_{\scr AB}$ built from the 5-dimensional
Riemann tensor $\mathcal{R}_{\scr ACBD}$.

The induced metric $h_{\scr AB}$ can be obtained
from the bulk metric $g_{\scr AB}$ together with
the inner unit vector $n^{\scr A}$ normal
to the brane through the relation $h_{\scr AB}
= g_{\scr AB} + n_{\scr A} 
n_{\scr B}$. Note that for a timelike extra-dimension
as we assume here, the vector $n^{\scr A}$ is
timelike, i.e. $g_{\scr AB} n^{\scr A} 
n^{\scr B} = -1$. In the metric \eqref{ds5} above,
its components read simply $n^{\scr A} = (1,0,0,0,0)$.

Assuming the brane's coordinates to be $\xi^\mu$, the 4-dimensional 
components of the induced metric, namely $h_{\mu\nu}$, are obtained by
\begin{align}
h_{\mu\nu} = g_{\scr AB}
\frac{\partial x^{\scr A}}{\partial \xi^\mu} 
\dfrac{\partial x^{\scr B}}{\partial \xi^\nu}.
\label{hmn}
\end{align}
In practice, one can set $\xi^\mu \to x^\mu$ for $\tau=0$,
so that Eq. \eqref{hmn} becomes merely the restriction to the
4-dimensional indices of $h_{\mu\nu}\to g_{\mu\nu}$.

In Eq.~\eqref{act5act4}, $K_{\scr AB}
= h^{\scr C}_{\ {\scr A}}
\nabla_{\scr C} n_{\scr B}$ is the
extrinsic curvature, $K=h_{\scr AB}
K^{\scr AB}$ representing its trace. Since
$K_{\scr AB}$ lies in the brane, i.e.,
$K_{\scr AB} n^{\scr A} =
K_{\scr AB} n^{\scr B} =0$,
one can restrict attention to $K_{\mu\nu} = h^\gamma_{\ \mu}
D_\gamma n_\nu$ ($D_\gamma$ denoting the covariant derivative
on the brane associated with the induced metric $h_{\mu \nu}$)
and obtain, from the variation of the action \eqref{act5act4}
with respect to $h_{\mu\nu}$, the 4-dimensional EFE
\begin{align}
m^2 G_{\mu \nu}+\sigma h_{\mu \nu}= T_{\mu \nu} -
M^3 \underbrace{\qty( K_{\mu \nu} - h_{\mu \nu} K )}_{\equiv S_{\mu\nu}},
\label{EFE51}
\end{align}
where $T_{\mu \nu}$ is the stress energy tensor obtained
from the brane lagrangian $\mathcal{L}$, $G_{\mu\nu}
= R_{\mu\nu} -\frac12 g_{\mu\nu} R$ the Einstein tensor
stemming from $h_{\mu\nu}$, and we have defined
$S_{\mu\nu} \equiv K_{\mu \nu} - h_{\mu \nu} K$. The 4- and
5-dimensional Riemann tensors are related via the Gauss
identity on the brane, i.e.,
\begin{align}
R_{\scr MNRS} =
h^{\scr A}_{\ {\scr M}}
h^{\scr B}_{\ {\scr N}}
h^{\scr C}_{\ {\scr R}}
h^{\scr D}_{\ {\scr S}}
\mathcal{R}_{\scr ABCD} + 
K_{\scr NR} K_{\scr MS} -
K_{\scr MR} K_{\scr NS},
\label{Gauss}
\end{align}
from which one gets $R_{\scr NS} = h^{\scr MR} R_{\scr MNRS}$, namely
\begin{align}
\begin{split}
R_{\scr NS} = & \mathcal{R}_{\scr NS} + n^{\scr A}
\qty(n_{\scr N} \mathcal{R}_{\scr AS} + n_{\scr S} \mathcal{R}_{\scr AN})
+ n^{\scr A} n^{\scr B} \mathcal{R}_{\scr ANBS} \\
 & + \qty(n^{\scr A} n^{\scr B} \mathcal{R}_{\scr AB})
n_{\scr N} n_{\scr S},
\end{split}
\label{RAB54}
\end{align}
where use has been made, because of \eqref{Gauss}, that the normal
$n_{\scr A}$ yields a vanishing contraction with the 4D Riemann tensor
with whatever component. 
Upon projection on the normal, Eq.~\eqref{EFE5} provides
$n^{\scr A} n^{\scr B} \mathcal{R}_{\scr AB} = \Lambda_5-\frac12
\mathcal{R}$, so that further contracting \eqref{RAB54} finally yields
\begin{align}
K_{\mu\nu} K^{\mu\nu} - K^2 = R - 2 \Lambda_5
=S_{\mu\nu}S^{\mu\nu}-\frac13 S^2.
\label{GaussScalar}
\end{align}
Using $S_{\mu\nu}$ from Eq.~\eqref{EFE51} and $G_{\mu\nu} h^{\mu\nu}
=- R$, we arrive at
\begin{align}
&(m^2G_{\mu\nu}+\sigma h_{\mu\nu}-T_{\mu\nu})
   (m^2G^{\mu\nu}+\sigma h^{\mu\nu} -T^{\mu\nu})
\nonumber\\
&-\tfrac{1}{3}(m^2R+T-4\sigma)^2 = M^6(R-2\Lambda_{5}) ,
\end{align}
where we have set $T\equiv h_{\mu\nu} T^{\mu\nu}$.



Defining $R_{\mu\nu} = h_\mu^{\scr N} h_\nu^{\scr S} R_{\scr NS}
$and using \eqref{EFE51}, one finds
\begin{align}
R_{\mu\nu} = h_{\mu}^{\scr N} h_{\nu}^{\scr S} \mathcal{R}_{\scr NS}
+ h_{\mu}^{\scr N} h_{\nu}^{\scr S} n^{\scr B} n^{\scr B}
\mathcal{R}_{\scr ANBS}
- K K_{\mu\nu} + K^{\alpha}_{\mu}K_{\nu \alpha}.
\end{align}
One can now decompose the 5D Riemann tensor into its Weyl and Ricci 
components as
\begin{align}
\begin{split}
\mathcal{R}_{\scr ANBS} & = W_{\scr ANBS} -\frac{\mathcal{R}}{12} 
\qty(g_{\scr AB} g_{\scr NS} - g_{\scr AS} g_{\scr NB})\\
& -\frac13 \qty(\mathcal{R}_{\scr AS} g_{\scr NB}
- \mathcal{R}_{\scr AB} g_{\scr NS} 
+ \mathcal{R}_{\scr NB} g_{\scr AS}
- \mathcal{R}_{\scr NS} g_{\scr AB})
\end{split}
\label{WR}
\end{align}
and define the symmetric and traceless projection of the bulk Weyl
tensor on the brane $W_{\mu\nu}\equiv h_{\mu}^{\scr N} h_{\nu}^{\scr S}
n^{\scr A} n^{\scr B} W_{\scr ANBS}$ to obtain the 4D Einstein tensor

Using the above result along with the 5D EFE on a non-vacuum bulk with stress-energy tensor $T_{AB}$ and decomposing the 5D Riemann tensor on the bulk into bulk Ricci tensor $\mathcal{R}_{AB}$, bulk scalar curvature $\mathcal{R}$ and the bulk Weyl tensor $W_{BDF}^A$, we get

\begin{equation}
\begin{aligned}
&G_{\mu\nu}= \frac{2}{3M^2} 
\Bigg[ 
T_{AB} h_\mu^{\,A} h_\nu^{\,B} 
+ \left( T_{AB} n^A n^B - \tfrac{1}{4} T \right) h_{\mu\nu} 
\Bigg] \\
&\quad + W_{\mu\nu} 
- K K_{\mu\nu} 
+ K_{\mu}^{\ \sigma} K_{\nu\sigma} 
- \tfrac{1}{2} h_{\mu\nu} \left( K^{\alpha\beta} K_{\alpha\beta}-K^2 \right),
\end{aligned}
\end{equation}

where $W_{\mu\nu}=W_{BDF}^An_{A}n^{D}h_{\mu}^Bh_{\nu}^F$ is the projection of the bulk Weyl tensor on the brane and is symmetric and traceless, mimicking additional radiation fluid on the brane.

Using an infinitesimal thin brane, Israel junction conditions $[h_{\mu\nu}]=0$ and $[K_{\mu\nu}]=\frac{1}{M^2}\bigg(S_{\mu\nu}-\frac{1}{3}S_{\mu\nu}\bigg)$, where in the natural choice of coordinates $x^A\equiv(x^{\mu},y)$ with $x^{\mu}\equiv(t,x^i)$ denoting the brane coordinates, $[\alpha]\equiv\lim_{y\to0^+} \alpha-\lim_{y\to0^-} \alpha$, assuming that the brane is located at $y=0$ we arrive at the final form of the modified EFE on the brane. Here, $S_{\mu\nu}$ turns out to be $-\sigma h_{\mu\nu}+T_{\mu\nu}$. The bulk energy momentum tensor is assumed to have the form $T_{AB}=\Lambda_{5} g_{AB}$ describing a de-Sitter bulk.

We finally end up with the following modified 4-dimensional EFE \cite{Sahni_2005}

\begin{align}\label{gauss}
G_{\mu \nu}+\Lambda_\text{eff}\, h_{\mu \nu}=8\pi G_\text{eff} \,
T_{\mu \nu}-\frac{1}{1+\beta}\qty(\frac{\Pi_{\mu \nu}}{M^6}-
W_{\mu \nu}),
\end{align}
in which we set
$
\beta = -\frac{2 \sigma m^2}{3M^6}, \quad 
\Lambda_\text{eff} = \frac{\Lambda_\textsc{rs}}{1+\beta}, \quad 
\Lambda_\textsc{rs} = \frac{\Lambda_5}{2} -
\dfrac{\sigma^2}{3M^6},
$
while the new effective rescaled gravitational coupling constant reads
$
8\pi G_{\text{eff}} = \frac{\beta}{m^2(1+\beta)}
= - \dfrac{2\sigma}{4M^6} \qty(1-\dfrac{2\sigma m^2}{4M^6})^{-1},
$
Following Randall-Sundrum (RS), we assume that the bulk cosmological 
constant and the term containing brane tension are fine-tuned such
that $\Lambda_\textsc{rs} = 0$. This requires, as already mentioned
in Eq. \eqref{ds5}, that the brane tension $\sigma$ must be negative.
This can also be seen in the above relation by demanding the effective
gravitational constant to be positive in the high energy limit for
which $\sigma m^2 \ll M^6$. Thus, the SS braneworld is characterized
by a negative tension brane and a positive bulk cosmological constant.


We define $E_{\mu \nu} \equiv m^2 G_{\mu \nu} - T_{\mu \nu}$,which is used to compute
\begin{align}
\Pi_{\mu \nu}=\frac{1}{3}EE_{\mu \nu}-E_{\mu \alpha}E^{\alpha}_{\nu}+\frac{1}{2}\bigg(E_{\alpha \gamma}E^{\alpha \gamma}-\frac{1}{3}E^2\bigg)h_{\mu \nu}~.
\end{align}
Here, $W_{\mu \nu}$ is the projection of the bulk Weyl tensor $W_{\mu \nu \alpha \gamma}$ on the brane defined as $W_{\mu \nu}=n^{\alpha} n^{\gamma} W_{\mu \alpha \nu \gamma}$. The correction terms are related to one another by the conservation equation
\begin{align}
D^{\mu} \qty(\Pi_{\mu \nu}-M^6 W_{\mu \nu})=0,	
\end{align}
where $D^{\mu}$ denotes the covariant derivative on the brane associated with the induced metric $h_{\mu \nu}$.

We first consider a static, spherically symmetric metric having a line element of the form 
\begin{align}\label{line element}
	\dd s^2=-e^{\nu(r)}\dd t^2+e^{\lambda(r)}\dd r^2+r^2\qty(\dd\theta^2+\sin^2 \theta \dd\phi^2).
\end{align}
We will first obtain the vacuum solutions of the modified EFE for this metric, which shall be required to describe exterior solutions to the homogeneous collapsing dust sphere.

For vacuum, we have $T_{\mu \nu}=0\;(p=\rho=0)$. Then the modified EFE on the brane of the form $G_{\mu \nu}=W_{\mu \nu}$, as $\epsilon=-1$ for a time-like extra dimension.  

From properties of algebraic symmetry, the projected Weyl tensor on the brane can in general be decomposed into the form
\begin{align}\label{wyel-tansor}
	W_{\mu \nu}=-\frac{1}{16\pi G_\text{eff}\sigma}\bigg[U(u_{\mu}u_{\nu}+\frac{1}{3}h_{\mu \nu})+P_{\mu\nu}+2\mathbb{Q}(_\mu u_\nu)\bigg],
\end{align}
such that the irreducible decomposition is done with respect to a 4-velocity field $u^{\mu}$ and has a metric projected orthogonal to it given by $h_{\mu \nu}=g_{\mu \nu}+u_{\mu}u_{\nu}$. Here, $U$ denotes the effective energy density on the brane arising from the bulk gravitational field. To ensure gravity is localized near the brane, $U$ must be a positive quantity. The term $P_{\mu \nu}$ represents the anisotropic stress of the effective matter on the brane, induced by the bulk gravitational field, and is responsible for generating an inherent pressure anisotropy. It is given by
$
P_{\mu \nu} = P \left[ r_{\mu} r_{\nu} - \frac{1}{3} h_{\mu \nu} \right],
$ where $r_{\mu}$ is a unit radial vector. The effective anisotropic stress on the brane due to bulk effects thus appears as a linear combination of $U$ and $P$. Finally, the vector $\mathbb{Q}_{\mu}$ represents an additional effective energy flux on the brane, giving rise to Coulomb-like or gravitomagnetic effects originating from the bulk gravitational field.

The modified vacuum EFE on the SS brane has the form
\begin{align}
    e^{-\lambda}\left(\frac{\lambda^\prime}{r}-\frac{1}{r^2}\right)+\frac{1}{r^2}
	&=-\frac{12U}{\rho_c},\label{eq8}\\
	e^{-\lambda}\left(\frac{\nu^\prime}{r}+\frac{1}{r^2}\right) -\frac{1}{r^2}
	&=-\frac{4U}{\rho_c}-\frac{8P}{\rho_c},\label{eq9}\\
	e^{-\lambda}\left(\frac{\nu''}{2}-\frac{\lambda^\prime \nu^\prime}{4}+\frac{{\nu^\prime}^2}{4}+\frac{\nu^\prime-\lambda^\prime}{2r}\right) &=  -\frac{4U}{\rho_c}+\frac{4P}{\rho_c}.\label{eq10}
\end{align}   
The bulk gravitational effects on the brane generate an inherent pressure anisotropy, as evident from Eqns. (\ref{line element}) and (\ref{wyel-tansor}). Here, the prime symbol ($'$) denotes differentiation with respect to the radial coordinate $r$. The parameter $\rho_c$ denotes the critical density, given by $\rho_c = 2|\sigma|$.

The vacuum is characterized by the fundamental property that it admits no preferred reference frame, implying that any reference frame can be considered comoving with respect to it. Under the assumption of spherical symmetry, the Petrov algebraic classification~\cite{Petrov}, which allows for an infinite number of comoving reference frames~\cite{Gliner}, can be employed to describe the vacuum. In this context, the energy-momentum tensor takes the canonical form
$T^{0}_{0} = T^{1}_{1},$ while spherical symmetry further implies $ T^{2}_{2} = T^{3}_{3}.$ In the braneworld scenario, these equalities between components of the stress-energy tensor are replaced by equalities among the effective stress-energy components. In the vacuum case, this effectively reduces to the equalities between components of the projected Weyl tensor on the brane:
$ W^{0}_{0} = W^{1}_{1}, \quad W^{2}_{2} = W^{3}_{3}.$
The effective stress-energy components on the brane, due to the projected Weyl tensor in the case of a SS brane, are given by,
\begin{align}\label{weyl}
	W^{\mu}_{\nu}&=\text{diag}\,\bigg(-\frac{12U}{\rho_c},-\frac{4U}{\rho_c}-\frac{8P}{\rho_c}, -\frac{4U}{\rho_c}+\frac{4P}{\rho_c},\nonumber\\&-\frac{4U}{\rho_c}+\frac{4P}{\rho_c}\bigg).
\end{align}
$W_{2}^{2}=W_{3}^{3}$ is automatically implied from the spherical symmetry consideration. On setting $W_{0}^{0}=W_{1}^{1}$, we have $\frac{12U}{\rho_c}=-\frac{4U}{\rho_c}-\frac{8P}{\rho_c}$. Now, as $W_{\mu \nu}$ acts as the effective stress-energy of some additional matter on the brane, transferring the bulk gravitational effects onto it, $U$ and $P$ are connected by the parameter $\omega$, such that $P=\omega U$. Plugging this in the above equality yields us $\omega=-2$. Thus, we have $U=-\frac{1}{2}P$. In turn, plugging the relation between $U$ and $P$ in the modified EFE Equations (4) and (5), we get the relation $\lambda'+\nu'=0$. Solving the differential equation, we get $\lambda(r)+\nu(r)=g(t)$, where $g(t)$ can be chosen to be zero without any loss of generality. This finally yields a relation between the temporal and radial metric tensor components of the form $\lambda(r)=-\nu(r)$.

Solving the modified field Eqn (\ref{eq8})-(\ref{eq10}) gives the metric potentials having the form
\begin{align}
	e^{\nu(r)}=r^{-\lambda(r)}=1-\frac{A}{r}+\frac{B}{r^2}.
\end{align}
The above-obtained solution closely resembles the Reissner–Nordström (RN) solution in GR. So, the two constant parameters $A$ and $B$ can be identified as $A=2 \mathbb{M}$ and $B=\mathcal{Q}$. However, since we have not considered any electric charge, the parameter $\mathcal{Q}$ represents the tidal charge which is physically responsible for the transmission of bulk gravitational degrees of freedom onto the brane via the projected bulk Weyl tensor. We also have
\begin{align}\label{weyl1}
W^{\mu}_{\nu}=\frac{\mathcal{Q}}{r^4}\text{diag}\,\qty(-1,-1,1,1).
\end{align}
In the case of the RS brane, there is a very interesting possibility that it is not permitted within standard GR premises, such as $\mathcal{Q}<0$. This leads to a very interesting case of RN black hole with a single horizon as for $\mathcal{Q}<0$, $g_{00}(r)$ has only one non-negative root. In the RS brane, it is justified that $\mathcal{Q}<0$ as the effective energy density $U$ on the brane arising from bulk contribution must be negative to keep gravity confined to the vicinity of the brane just like the contribution of negative $\Lambda_{5}$ in causing an acceleration towards the brane. On the contrary, for SS brane $\Lambda_{5}$ is positive, and for the confinement of gravity in the vicinity of the negative tension brane, the energy density $U$ must be positive. It does not need to be negative as we do not consider any form of bulk matter. Thus, comparing Eqns. (\ref{weyl}) and (\ref{weyl1}), we see that $\mathcal{Q}>0$ is used for the SS-brane. Thus, the bulk effect increases the strength of the gravitational field in the RS model. At the same time, tidal charge weakens the gravitational field in the SS model, similar to the weakening effect of electric charge in RN black holes in GR. However, there is a key distinction between SS gravity and both GR and RS gravity in the nature of the singularity: it is time-like in the SS branch, whereas it is space-like in GR and RS gravity.  It is evident that, although the strong energy condition (SEC) is marginally satisfied by the effective matter on the brane, as indicated by the relation $\rho^{\text{eff}} + p_r^{\text{eff}} + 2p_t^{\text{eff}} = 0$, the null energy condition (NEC) is violated in the radial direction as well as on average. In addition, both the weak and dominant energy conditions are also violated. Here $\rho^{eff}$, $p_{r}^{eff}$, and $p_t^{eff}$ are the effective energy density, effective radial pressure, and effective transitional pressure on the brane, respectively.  These violations suggest the possibility of avoiding the formation of a singularity.

\section{Gravitational collapse} \label{collapse}

\subsection{Exterior}

Assuming a general form of static, spherical symmetric metric on the brane to describe the exterior of a collapsing mass, having the form
\begin{align}
	\dd s^2=-F(r)\dd t^2+ \frac{\dd r^2}{F(r)}+ r^2\dd\Omega^2,
\end{align}
where $F(r)=1-2 \mathbb{M}(r)/r$, the Ricci scalar (in absence of a $\Lambda$-term) turns out to have the form
\begin{align}
	R^{\mu}_{\mu}=-\frac{9 \mathbb{M}^2}{2\pi\sigma r^6}.
\end{align}
However, the Ricci scalar for the vacuum exterior of the collapsing mass on the brane in the absence of $\Lambda$ should vanish as implied by Eq. (\ref{gauss}). This automatically implies $\mathbb{M}/\sigma=0$ for Birkhoff's theorem to hold true that the spherically symmetric exterior vacuum solution must be static. This means either the mass must vanish which is physically meaningless or the brane tension must be infinitely large (which is true in standard GR). So, Birkhoff's theorem is violated on the brane but can be recovered in the GR setting. Hence, we cannot describe the exterior solution on the brane using a static vacuum, and it has to be generalized.

The simplest possible spherically symmetric generalization of the static vacuum is the Vaidya spacetime that is used to describe radiating mass in GR. It describes the simplest non-static exterior and has a non-vanishing $T_{\mu \nu}$ as the energy density of the radiation is accounted for. The stress-energy tensor has the form given by $T_{\mu \nu}=\alpha k_{\mu}k_{\nu}$, such that $k_{\mu}k^{\mu}=0$ and $\alpha$ represents the energy density of the radiation or null fluid. For the null fluid/radiation on the brane, we have $S_{\mu\nu}=0$. The modified EFE takes the form
\begin{align}
	G_{\mu\nu}=\alpha k_{\mu}k_{\nu}+W_{\mu \nu}.
\end{align} 
The solution is best expressed in the Eddington-Finkelstein coordinates and has the form
\begin{align}\label{Edd-Fink}
ds^2 = -\left( 1 - \frac{2 \bar{m}(v)}{r} + \frac{q(v)}{r^2} \right) dv^2 + 2\, dv\, dr + r^2 d\Omega^2 .
\end{align}
where $v$ denotes the advanced time, related to the previous variables through 
the transformation
$$
\dd v=\dd t+ \qty(1-\frac{2\mathbb{M}}{r}+\frac{\mathcal{Q}}{r^2})^{-1}
\dd r.
$$
The standard relativistic result is that a collapsing sphere does not radiate, but in the braneworld scenario, in order to accommodate the non-static nature of the exterior solution, the simplest possible generalization is to include matter in the form of null fluid, thus leading to radiation from the collapsing sphere, as obtained earlier in the case of RS brane \cite{ND1}. The exterior thus comprises of a Vaidya radiation envelope, which is finally matched to the static vacuum described by the RN solution. The continuous matching between the interior and the static vacuum can not be done due to a violation of Birkhoff's theorem on the brane. 
 
\subsection{Interior} 

The interior of the collapsing dust cloud on the brane can be described by the Friedmann-Lemaitre spacetime written as
\begin{align}\label{fried}
	\dd s^2=-\dd\tau^2+a(\tau)^2\qty(\frac{\dd r^2}{1-kr^2}+r^2\dd\Omega^2),
\end{align}
For our purpose, it is convenient to rewrite the above line element with spherical spatial sections and we consider a spatially closed geometry with positive spatial curvature $k=1$, yielding the form
\begin{align}\label{positive}
	\dd s^2=-\dd\tau^2+a(\tau)^2\qty(\dd\chi^2 + \sin^2 \chi \dd\Omega^2),
\end{align}
where $R=a(\tau)r=a(\tau) \sin \chi$.  
 
\subsection{Mathematical model of collapse}

We get the first junction condition by ensuring smooth matching of the metric functions at the surface for the metrics given in Eqns. (\ref{Edd-Fink}) and (\ref{fried}). This leads us to the relation
\begin{align}
\dd\tau = \dd v\, \sqrt{1-\frac{2\bar{m}(v)}{r}+\frac{q(v)}{r^2}-2\frac{\dd r}{\dd v}}~.
\end{align}
This is almost identical to the similar relation obtained for the RS brane.

Also, the induced metrics on the boundary surface must match smoothly which can be done  by matching Eq. (\ref{positive}) to the induced exterior metric on the surface, written in the form 
\begin{align}
\dd s^2 = - \qty[g \qty(\frac{\dd v}{\dd \tau})^2-\frac{1}{g}\qty(\frac{\dd R}{\dd\tau})^2]\dd\tau^2 + r^2 \dd\Omega^2,
\end{align} 
where 
$$
g=1-\frac{2 \bar{m}(v)}{r}+\frac{q(v)}{r^2} -2\frac{\dd r}{\dd v}.
$$

The junction condition yields
\begin{align}
g \qty(\frac{\dd v}{\dd\tau})^2 - \frac{1}{g} \qty( \frac{\dd R}{\dd\tau})^2
=1 ,
\end{align}
which may equivalently be written as
\begin{align}
1 - \frac{2 \bar{m}(v)}{R} + \frac{q(v)}{R^2} -2 \frac{\dd R}{\dd v} =
\qty(\frac{\dd R}{\dd\tau})^2+ g.
\end{align}
After some simplification, one gets 
\begin{align}
v = - 3 \int \qty[1-\frac{2 \bar{m}(v)}{R}+\frac{q(v)}{R^2}]^{-1} \dd R.
\label{e8}
\end{align}
Here, it is to be noted that we take the negative root of $\dd R/\dd\tau$ as it implies the case for collapse. The positive root would have implied expansion, which we shall find to be accounted for later.

The second junction condition involves continuity of the extrinsic curvature $K_{\mu \nu}$ which equivalently involves the continuity of $\dd R/\dd\tau$, and it leads us to a relation for the dynamic mass of the form
\begin{align}\label{mass}
	\bar{m}(v)=\bar{m}-\frac{3\bar{m}^2}{2\sigma R^3}-\frac{\mathcal{Q}-q(v)}{2R\sigma}.
\end{align}
Since we are considering the collapse of a homogeneous dust cloud, the pressure $p=0$ for dust matter and this leads us to a relation for the dynamic tidal charge at the surface, given as
\begin{align}\label{charge}
	q(v)=\qty(\frac{3\bar{m}^2}{R^2}+\mathcal{Q})\bigg|_\text{surface}.
\end{align} 
Plugging in Eq. (\ref{charge}) into Eq. (\ref{mass}), we have $\bar{m}(v)=\bar{m}$. This is identical to the condition obtained for Oppenheimer-Snyder Collapse in standard GR.

However, the matching exercise cannot be completed without matching the nonstatic Vaidya radiation envelope to the static exterior, which is also modified to RN spacetime due to the presence of tidal charge arising from the non-local stresses being propagated from the bulk via the projected Weyl tensor on the brane. This finally leaves us with $\bar{m}(v)=\mathbb{M}$ and $q(v)=\mathcal{Q}$.  Putting these values in Eq. (\ref{e8}), we eventually get
\begin{eqnarray}\label{e9}
v&=&-\frac{3}{\sqrt{\mathbb{M}^2-\mathcal{Q}}}\Big\{ \qty( 2 \mathbb{M}^2-\mathcal{Q}) \arctanh \qty( 
\frac{\mathbb{M}-R}{\sqrt{\mathbb{M}^2-\mathcal{Q}}})\nonumber\\
& &+ \qty[ \mathbb{M}\ln  \qty( 2\,\mathbb{M} R-{R}^{2}-\mathcal{Q}) + R]\sqrt {\mathbb{M}^{2}-\mathcal{Q}} \Big\}.
\end{eqnarray}
As $R_{\pm}=\mathbb{M}\pm\sqrt{\mathbb{M}^2-\mathcal{Q}}$ at the horizon, we have $v\rightarrow \infty$. So, the process of forming a horizon is infinitely long. 
\begin{figure}
\begin{center}
\includegraphics[width=0.38\textwidth]{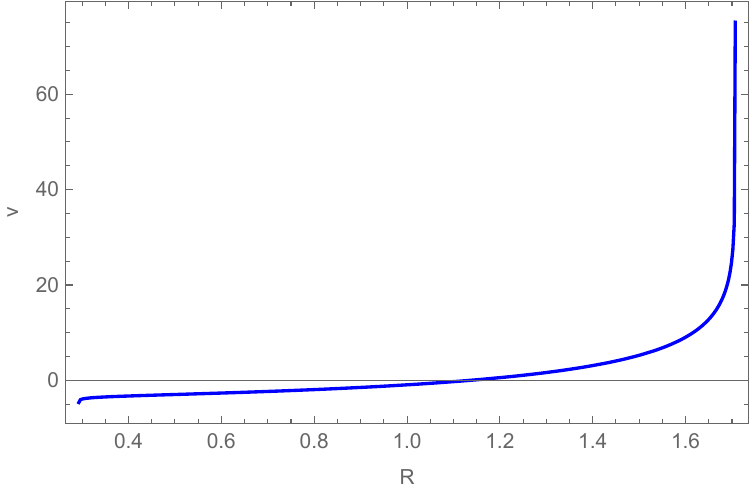}
\caption{Plot of $v$ as a function of $R$}
\label{apparent_con}
\end{center}
\end{figure}  

The $v$-$R$ plot (Figure  \ref{apparent_con}) shows that as the collapse progresses, the advanced time $v$ diverges at a finite radius ($R \simeq 1.6$), signaling the formation of a horizon. Beyond this point, outgoing null rays are infinitely delayed, indicating the trapping of light. In our model, however, the interior evolution avoids a singularity: the collapse can either undergo a bounce, leading to re-expansion, or settle into a stable remnant core. Thus, the figure captures horizon formation without singularity, distinguishing our scenario from classical black hole collapse.

For the interior of the collapsing dust cloud described by Eq. (\ref{fried}), the modified EFE on the brane for $\epsilon=-1$ and a positive spatial curvature $k=+1$ yields the Friedmann equation
\begin{align}\label{fried1}
\qty( \frac{\dot{a}}{a})^2 + \frac{1}{a^2} = 
\frac{8\pi}{3}\qty(\rho-\frac{\rho^2}{2\sigma}) + \frac{C}{a^4},
\end{align}
where we can set the integration constant $C=0$ for simplicity.
From the above equation, it follows that as the energy density increases due to the decreasing volume of the collapsing matter, the negative quadratic energy density term begins to dominate. In the limit $\rho \rightarrow 2\sigma$, one finds that $\frac{\dot{a}}{a} + \frac{1}{a^2} \rightarrow 0$. Consequently, the energy density and the scalar curvature remain finite due to the braneworld corrections, and the formation of a singularity is thus avoided. In other words, the collapse is halted as a result of extra-dimensional effects. While the post-collapse dynamics will be discussed in due course, at this stage we aim to determine the minimum value of the scale factor and the corresponding critical energy density of the collapsing dust sphere, defined as the scale factor and energy density, respectively, for which $\dot{a} = 0$.
Plugging in $\dot{a}=0$ in Eq. (\ref{fried1}), we get
\begin{align}
\frac{1}{a_\text{min}^2}=\frac{8\pi}{3}\qty( \rho_\text{cr} -
\frac{\rho_\text{cr}^2}{2\sigma}),
\end{align}
where the critical density is obtained as
\begin{eqnarray}\label{E1}
\rho_\text{cr}=\sigma\pm\sqrt{\sigma^2 -\frac{3 \sigma}{4\pi  a_\text{min}^2}}.
\end{eqnarray}
The term under the square root must be non-negative, and at $a=a_\text{min}$, it vanishes. So, we have 
\begin{eqnarray}\label{e2}
a_\text{min} = \sqrt{\frac{3}{4 \pi\sigma}} \quad \hbox{and} \quad
\rho_\text{cr}=\sigma.
\end{eqnarray}
Thus, the density at which the collapse stops due to braneworld corrections is equal to the brane tension, and the corresponding scale factor is also a function of the brane tension, such that $a_\text{min}^2 \propto 1/\sigma$.

Let us now consider the scenario at the moment the collapse starts, denoted by time $\tau=0$. At $\tau=0$, the scale factor is at its maximum, as the dust cloud has not yet started collapsing. The maximum scale factor is denoted by $a_0$. We consider 
\begin{align}
	a_0=\frac{8\pi \rho a^3}{3}.
\end{align} 
For pressureless dust matter ($p=0$), the conservation equation on the brane $\dot{\rho}+3\frac{\dot{a}}{a}(\rho+p)=0$ yields $\rho=\rho_0/a^3$. So, plugging the $\rho$ in Eq. (31), it turns out that $a_0$ is a constant, as expected. The corresponding radius of the dust cloud at $\tau=0$ is given by $R(\tau=0)=a(\tau=0)\sin \chi_{(\tau=0)}=a_0 \sin \chi_{0}=R_0$. Using this relation it is possible to define $a_0$ and $\sin \chi_{0}$ in the following form, where $R_{\pm}=\mathbb{M}\pm\sqrt{\mathbb{M}^2-\mathcal{Q}}$ is the horizon corresponding to the vanishing roots of $g_{00}$, which we have found takes infinitely long time to form physically, owing to braneworld corrections. We have
\begin{eqnarray}\label{E23}
a_0&=&\sqrt{\frac{R_{0}^3}{\mathbb{M}\pm\sqrt{\mathbb{M}^2-\mathcal{Q}}}}\nonumber\\
\sin \chi_{0}&=&\sqrt{\frac{\mathbb{M}\pm\sqrt{\mathbb{M}^2-\mathcal{Q}}}{R_0}}.
\end{eqnarray}
Eq. (\ref{fried1}) for $k=1$ and $C=0$ can be rewritten in the form 
\begin{eqnarray}\label{e10}
{\dot{a}(\tau)}^2+1=\frac{a_0}{a(\tau)}\left(1-\frac{\rho_0}{\rho_c a(\tau)^3}\right).
\end{eqnarray}
We further rewrite this equation in terms of the conformal time $\eta$, such that $d\tau=a(\eta) d(\eta)$
\begin{eqnarray}\label{e100}
	{\dot{a}(\eta)}^2+a(\eta)^2=a_0{a(\eta)}\left(1-\frac{\rho_0}{\rho_c a(\eta)^3}\right).
\end{eqnarray}
\begin{figure}
\begin{center}
\includegraphics[width=0.38\textwidth]{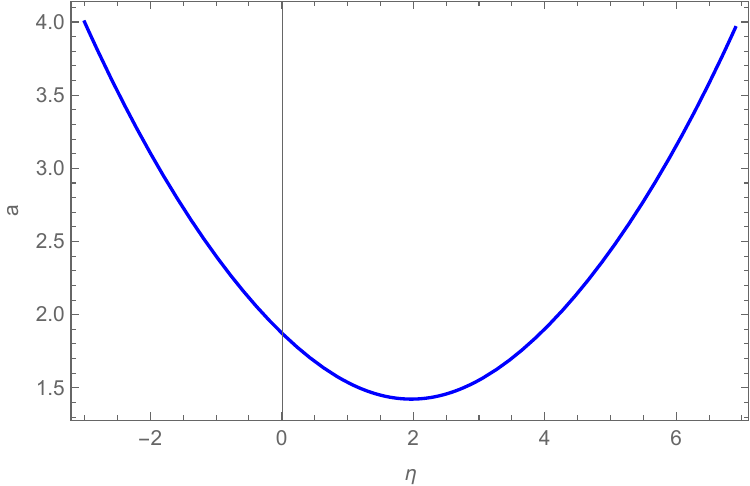}
\caption{Variation of $a$ with respect to $\eta$}
\label{apparent_con3}
\end{center}
\end{figure}

It is not possible to solve the above differential equation analytically and we resort to choosing the parametric form
\begin{eqnarray}\label{E21}
	a(\eta)=\frac{a_0}{2}\qty[1+\xi(\eta)],
\end{eqnarray}
where 
\begin{eqnarray}\label{98}
	\xi \left( \eta \right) &=&\pm \int {\frac{2\sqrt {-\sigma \left(  \left( a \left( \eta \right)  \right)^{4}\sigma - a_{{0}}\sigma \left( a \left( \eta \right)  \right) ^{3}+a_{{0}}\rho_{{0}} \right) }}{\sigma a \left( \eta \right) a_{{0}}}}
	{\rm d}\eta\nonumber\\ & & +C_3~.
\end{eqnarray}
The two signs `+' and `-' correspond to two branches of the solution admitting expansion and contraction. $C_3$ denotes the constant of integration.

Figure \ref{apparent_con3} shows the evolution of the scale factor a($\eta$) of a collapsing dust cloud on the brane, plotted against conformal time $\eta$. Physically, it illustrates a non-singular gravitational collapse scenario: the dust cloud initially contracts under gravity, with a($\eta$) decreasing as $\eta$ increases, reaching a minimum non-zero value instead of collapsing to a singularity. This minimum represents a bounce caused by high-energy brane-world corrections, which introduce repulsive effects at small scales, halting the collapse. After the bounce, the cloud re-expands symmetrically, as seen in the increasing a($\eta$) for larger $\eta$. The plot thus captures a complete collapse-bounce-expansion cycle, showing how brane-world gravity avoids singularities that are inevitable in classical General Relativity.

We consider the time at which collapse halts due to the braneworld corrections to be denoted by $\tau_\text{max}$. At this time $\tau=\tau_\text{max}$, the scale factor must have its minimum value, which is non-zero, denoted by $a=a(\tau_\text{max})=a_\text{min}$. Since the mass is conserved during the collapse process, we shall have 
\begin{align}
	a(\tau_\text{max})=\frac{8\pi \rho a^3}{3}, 
\end{align} 
which shall be a constant as discussed earlier. Similar to the situation at $\tau=0$, we have $R_{\tau_\text{max}}=a(\tau_\text{max}) \sin \chi_{\tau_\text{max}}=a_\text{min} \sin \chi_{\tau_\text{max}}=R_{\tau_\text{max}}$. So, in a similar way, we can write
\begin{eqnarray}\label{E231}
a(\tau_\text{max})=\sqrt{\frac{R_{\tau_{max}}^3}{\mathbb{M}\pm\sqrt{\mathbb{M}^2-\mathcal{Q}}}} \nonumber \\ \text{and} \ \ \ \ \ \sin \chi_{\tau_\text{max}}=\sqrt{\frac{\mathbb{M}\pm\sqrt{\mathbb{M}^2-\mathcal{Q}}}{R_{\tau_\text{max}}}}~.
\end{eqnarray}
The radius of the collapsed dust cloud at which the collapse halts can be calculated using the relation
\begin{eqnarray}\label{e11}
R_s|_{\tau_\text{max},a_\text{min}}={a(\tau_\text{max})}\sin\chi_\text{max}.
\end{eqnarray}
So, using Eqns. (\ref{e2}) and (\ref{E231}), we finally have
\begin{eqnarray}\label{E26}
		R_s&=&R|_{\tau_\text{max},a_\text{min}}=a_\text{min} \sin \chi_{\tau_\text{max}}\nonumber\\
		&=&a_\text{min}\bigg(\mathbb{M}\pm\sqrt{\mathbb{M}^2-\mathcal{Q}}\bigg)^{\frac{1}{2}}R_{\tau_\text{max}}^{-\frac{1}{2}}\nonumber\\
		&=&\bigg(\frac{4\pi \sigma}{3}\bigg)^{-\frac{1}{3}}  \bigg(\mathbb{M}\pm\sqrt{\mathbb{M}^2-\mathcal{Q}}\bigg)^{\frac{1}{3}}.
\end{eqnarray}
We find that the radius depends on the mass, tidal charge, and the brane tension.

The process of the collapse of the homogeneous dust cloud can also be viewed from the point of view of a free-falling observer. This mostly depends on the geometry of the exterior spacetime of the collapsing sphere. First, we need to compute the minimum radius $r_\text{m}$ of the collapse which can be obtained using the relation
\begin{align}
	\frac{\dd f}{\dd r}\bigg|_{r=r_\text{m}}=0,
\end{align}
where $f(r)=|g_{00}|=1-{\frac{2\mathbb{M}}{r}}+\frac{\mathcal{Q}}{r^2}$. This gives us 
\begin{eqnarray}\label{e16}
	r_\text{m}=\frac {\mathcal{Q}}{\mathbb{M}}.
\end{eqnarray}
If the minimum radius must be identical to the minimum scale factor which gives us a relation between the mass and tidal charge in terms of the brane tension of the form 
\begin{eqnarray}\label{e17}
	\mathcal{Q}=\frac{\mathbb{M}}{2} \sqrt{{\frac {3}{\pi \sigma}}}.
\end{eqnarray}
The total collapse time measured by the free-falling observer can be calculated to have the form
\begin{eqnarray}\label{e13}
T&=&\int_{0}^{T}\dd t=-\int_{r_+}^{r_\text{m}}\dd r\nonumber\\
&=& \mathbb{M}\pm\sqrt{\mathbb{M}^2-\mathcal{Q}}-{\frac{\mathcal{Q}}{\mathbb{M}}}.
\end{eqnarray}
Since both the mass and tidal charge parameters are finite, so the proper time also turns out to be finite for a freely falling observer.  Similarly, it will take a finite time for a light signal to reach the minimal distance from the outer horizon and turn out to have the form
\begin{eqnarray}\label{e14}
T_{0}&=&\int_{0}^{T_{0}}\dd t=-\int_{r_+}^{r_\text{m}}\frac{1}{2}\dd r\nonumber\\
&=&\frac{\mathbb{M}}{2}\pm\frac{1}{2}\sqrt{\mathbb{M}^2-\mathcal{Q}}-{\frac {\mathcal{Q}}{2 \mathbb{M}}}.
\end{eqnarray}
The time of collapse as measured by a faraway observer from the collapsing dust sphere can be computed as
\begin{eqnarray}\label{e15}
t&=&T-\int \frac{\dd r}{f(r)}\nonumber\\
t&=&T-r-\mathbb{M}\ln  \left( 2 \mathbb{M} r-{r}^{2}-\mathcal{Q} \right) \nonumber\\&&-2{\frac {\mathbb{M}^{2}}{\sqrt{\mathbb{M}^2-\mathcal{Q}}}{\rm arctanh} \left({\frac{\mathbb{M}-r}{\sqrt{\mathbb{M}^2-\mathcal{Q}}}}\right)}\nonumber\\
&&+{\frac {\mathcal{Q}}{\sqrt{\mathbb{M}^2-\mathcal{Q}}}{\rm arctanh} \left({\frac {\mathbb{M}-r}{\sqrt{\mathbb{M}^2-\mathcal{Q}}}}\right)}.
\end{eqnarray}
\begin{figure}
\begin{center}
\includegraphics[width=0.38\textwidth]{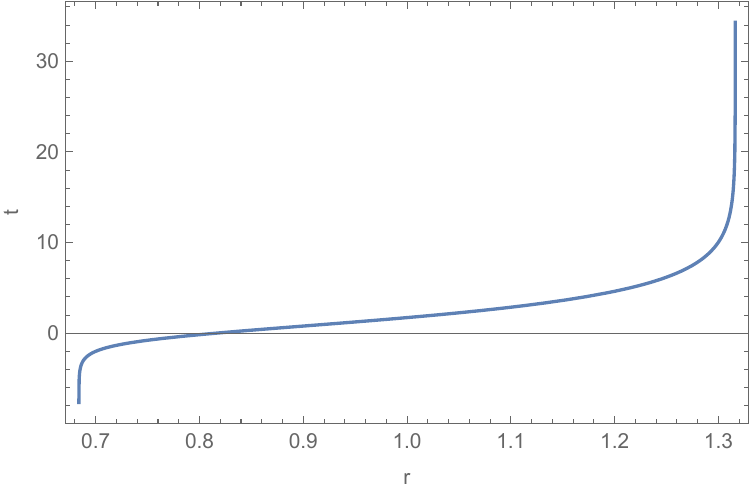}
\caption{Plot showing the variation of $t$ with respect to $r$.}
\label{apparent_con1}
\end{center}
\end{figure}  
In this case, once again we see that the time taken for the collapsing dust cloud to reach the horizon at $r=\mathbb{M}\pm\sqrt{\mathbb{M}^2-\mathcal{Q}}$ takes an infinitely long time as measured by a distant observer. The Ricci scalar $\mathcal{R}$ can be computed using the relation
\begin{eqnarray}\label{e20}
	\mathcal{R}&=&8 \pi \rho_\text{crit}=8 \pi \qty(\sigma\pm\sqrt{\sigma^2 -\frac{3k \sigma^2}{4\pi G a_\text{min}^2}})\nonumber\\&=&8 \pi\sigma.
\end{eqnarray}
As we see the scalar curvature depends only on the brane tension. So, for the scalar curvature to diverge, the brane tension must also diverge, which is true only when we shift over to standard GR, and there will be a formation of a singularity. However, as long as braneworld effects are significant, the brane tension remains finite and so does the scalar curvature.   

As shown in Figure \ref{apparent_con1}, the collapse time 
$t$ increases sharply as the radial coordinate 
$r$ approaches a finite value, beyond which it diverges. This behaviour signifies the formation of a trapping surface, as seen by an external observer. In contrast to the classical general relativistic case, where such divergence accompanies the development of a curvature singularity, our model ensures a regular interior evolution. The collapse halts at a minimum radius, avoiding singularity formation and opening the possibility of a subsequent bounce.

\section{Post collapse dynamics}\label{collapsehalts}

\subsection{Case - 1}

There are two possibilities: First, a non-singular black hole is formed with a dense core inside which the density and curvature values remain finite. There are two horizons: $r_{+}$ is the outer horizon, and $r_{-}$ is the inner horizon. Due to the presence of a horizon, Hawking radiation is taken into consideration. The outer horizon turns out to have the form
\begin{align}
	r_{+} \approx 2\mathbb{M}-\frac{\mathcal{Q}}{2\mathbb{M}}.	
\end{align}
For black holes of large mass, the outer horizon is close to the Schwarzschild radius. The Hawking temperature of the regular BH formed can be computed using
\begin{eqnarray}\label{hawking}
	T_\textsc{h}&=&\frac{1}{4\pi}\frac{\dd f}{\dd r}\bigg|_{r=r_{+}}\nonumber\\ &=&{\frac{1}{4\pi} \qty[ {\frac{2\mathbb{M}}{\left( 2\mathbb{M}-\frac{\mathcal{Q}}{2\mathbb{M}} \right)^{2}}}-{\frac{2\mathcal{Q}}{ \left( 2\mathbb{M}-\frac{\mathcal{Q}}{2\mathbb{M}} \right)^{3}}} ] }.
\end{eqnarray}
Due to the emission of Hawking radiation from the horizon, there is a consequent decrease in mass of the collapsed spherical cloud of dust, the rate of which can be expressed as
\begin{eqnarray} \label{evaporation_time}
	&-\displaystyle\frac{\dd \mathbb{M}}{\dd t}& \approx A\beta {T_{H}}^4= 4 \pi {r_{+}}^2 \beta {T_{H}}^4\nonumber\\&=& \beta\bigg(\frac{1}{4\pi}\bigg)^3\left[{\frac{2\mathbb{M}}{\left( 2\mathbb{M}-\frac{\mathcal{Q}}{2\mathbb{M}} \right)^{\frac{3}{2}}}}-{\frac{2\mathcal{Q}}{ \left( 2\mathbb{M}-\frac{\mathcal{Q}}{2\mathbb{M}} \right)^{\frac{5}{2}}}}\right]^{4},\nonumber\\
\end{eqnarray}
where $\beta$ denotes the Stefan-Boltzmann constant.

From Eq. (\ref{evaporation_time}), we obtain the BH evaporation time as, 
\begin{align}\label{BHEva}
	t_{e} \approx c_{1}\left(\mathbb{M}^3-\mathbb{M}_{e}^3\right) + \frac{c_{2}}{\sigma}\left(\mathbb{M}-\mathbb{M}_{e}\right),
\end{align}
where $c_1$ and $c_2$ are arbitrary constants, $\mathbb{M}_{e}=\sqrt{Q}$ denotes an extremal mass below which no real root of $g_{00}$ can be found. Also, at $\mathbb{M}=\mathbb{M}_{e}$, the two horizons $r_{+}=r_{-}=\sqrt{3/4\pi\sigma}=a_{min}$. Due to the braneworld effects, the black hole evaporation time increases. Another significant difference from GR is that in GR, the temperature increases, and the BH gets hotter, but here, due to the presence of a tidal charge, it gets cooler before vanishing at $\mathbb{M}=\mathbb{M}_{e}$. This can be checked in the following way. If we consider the mass to be extremal in Eq. (\ref{e17}), then $\mathbb{M}_{e}=\sqrt{3/4\pi\sigma}=r_{+}$. Plugging this in Eq. (\ref{hawking}), we have $T_{H}=0$. Interestingly, the outer horizon and the inner horizon coincide at the minimal scale factor, as the collapsing mass can't contract further. We are left with a remnant core, which, if stable, can act as a possible candidate for dark matter.

\subsection{Case - 2}

However, if the internal core formed inside the horizon is unstable, then we have a second possibility: once the collapse is halted, a bounce occurs, and the dust cloud starts expanding. This is where the positive branch of Eq. (\ref{98}) comes into play. This possibility has no analogue in GR. The bounce takes place at $a=a_\text{min}$, and the ratio $\dot{a}/a$ vanishes at the bounce. Also $\ddot{a}>0$ at the bounce.  The second modified Friedmann equation on the SS brane has the form,
\begin{eqnarray}\label{E27}
	{\frac{\ddot{a}}{a}} =-\frac{4 \pi}{3} \qty[ \rho+3p-{\frac {\rho \qty( 2\rho+3p ) }{\sigma}} ].
\end{eqnarray}
Let us generalize the scenario to include any type of fluid such that the Equation of state (EOS) is given as $p_\text{cr}=w \rho_\text{cr}$. We need to consider the critical density as that is the density of the fluid at the bounce. If we plug the above relation into Eq. (\ref{E27}), the equation becomes extremely simple and may be written as
\begin{align}\label{e19}
\frac{\ddot{a}}{a}=-\frac{4\pi}{3}\sigma.
\end{align}
It is worth noting that although we have generalized the EoS to consider the collapse of any type of fluid, it turns out that the evolution of the scale factor at the bounce is independent of the EoS parameter $w$, i.e., the nature of the fluid. This is reminiscent of the fluid-independent nature of the cosmological bounce one obtains on the SS brane \cite{Sahni4}. It is an easy task to solve Eq. (\ref{E27}) and the general solution has the form
\begin{align}\label{E24}
a(\tau)=A_1 \exp(p \tau)+A_2 \exp(-p \tau).
\end{align}
Here $p=\sqrt{4 \pi \sigma/3}$. The bounce occurs at the minimum scale factor. So, we have $a(\tau=0)=a_\text{min}$ now at the bounce. This initial condition can be satisfied if we set $A_1=a_\text{min}$ and $A_2=0$. Thus, we have the solution 
\begin{eqnarray}\label{E25}
a(\tau)=a_\text{min} \exp(p \tau).
\end{eqnarray}

\begin{figure}
\begin{center}
\includegraphics[width=0.38\textwidth]{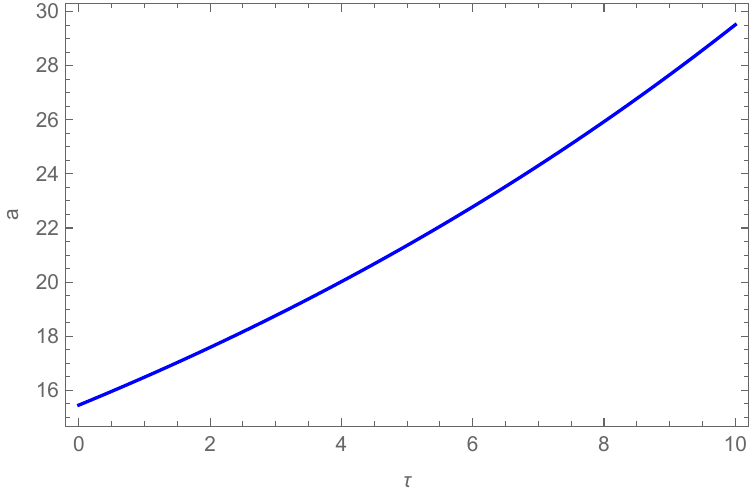}
\caption{Variation of $a$ with respect to $\tau$}
\label{apparent_con4}
\end{center}
\end{figure} 

The immediate post-bounce behavior shown in Figure \ref{apparent_con4} corresponds to the region just after the scale factor reaches its minimum. Physically, this marks the transition from collapse to expansion: the inward motion of the dust cloud halts, and it begins to re-expand. Right after the bounce, $a$ starts increasing slowly, with a small but positive time derivative $\dot{a}(\tau)$, indicating a gradual outward motion. This slow initial expansion is driven by the repulsive brane-world corrections that dominated near the bounce, overcoming the attractive gravitational pull. As time progresses further from the bounce, these corrections weaken, and the expansion rate is modified in accordance with the modified Friedmann dynamics.

Thus we find a very interesting result that the homogeneous dust cloud (or any fluid as there is no EoS dependence) after collapsing to a minimal scale factor bounces and following the bounce there is an exponential expansion of the deSitter type where the absence of the cosmological constant is made up for by the brane tension $\sigma$.
\noindent The interior metric looks like
\begin{align}
	\dd s^2 = - \dd\tau^2 + a_\text{min}^2 e^{2\sqrt{\frac{16\pi\sigma}{3}}t}
    \qty(\dd\chi^2+\sin^2 \chi \dd\Omega^2).
\end{align}
We also have
\begin{align}
	R(\tau)=a(\tau) \sin \chi=a_\text{min} e^{p\tau}\sqrt{\frac{\mathbb{M}\pm\sqrt{\mathbb{M}^2-\mathcal{Q}}}{R_s}}.
\end{align}
If we put $\tau=0$ in the above equation, then it turns out that
\begin{align}\label{e65}
R(\tau=0)=\bigg({\frac{3}{4\pi \sigma}}\bigg)^{\frac{1}{3}}\bigg(\mathbb{M}\pm\sqrt{\mathbb{M}^2-\mathcal{Q}}\bigg)^{\frac{1}{3}}=R_{s}.
\end{align}
So, we obtain the result that the expansion begins at the radius of the collapsed dust cloud that we had obtained earlier in Eq. (\ref{E26}), as expected. Once the exponential expansion begins, the expanding dust cloud will cross the horizon in a finite time. Let us consider this horizon crossing time as $\tau=\tau_{c}$. So, at $\tau=\tau_{c}$, we shall have $R(\tau=\tau_{c})=\mathbb{M}\pm\sqrt{\mathbb{M}^2-\mathcal{Q}}$. Plugging this result into Eq. (\ref{e65}), we obtain the shell crossing time as 
\begin{eqnarray}\label{E28}
\tau_{c}=\sqrt{\frac{3}{16\pi \sigma}}\ln\left[\frac{(\mathbb{M}\pm\sqrt{\mathbb{M}^2-\mathcal{Q}})R_{s}}{a_\text{min}^2}\right].
\end{eqnarray}
However we can also obtain the time taken to cross the shell from the point of view of an outside observer. Since we are considering an expansion of the dust cloud, so, we take the positive root of $\dd R/\dd\tau$ in this case. The time for the outside observer watching the expansion process far away from the expanding dust cloud turns out to be
 \begin{eqnarray}\label{981}
	t&=&R+\mathbb{M}\ln  \left( 2\mathbb{M}R-{R}^{2}-\mathcal{Q} \right)\nonumber\\
	&& +2{\frac {\mathbb{M}^{2}}{\sqrt{\mathbb{M}^2-\mathcal{Q}}}{\rm arctanh} \left(1/2{\frac {2\mathbb{M}-2R}{\sqrt{\mathbb{M}^2-\mathcal{Q}}}}\right)}-\nonumber\\
	&&{\frac {\mathcal{Q}}{\sqrt{\mathbb{M}^2-\mathcal{Q}}}{\rm arctanh} \left(1/2{\frac {2\mathbb{M}-2R}{\sqrt{\mathbb{M}^2-\mathcal{Q}}}}\right)}-R_{{s}}\nonumber\\
	&&-\mathbb{M}\ln  \left( 2\mathbb{M}R_{{s}}-{R_{{s}}}^{2}-\mathcal{Q}\right) \nonumber\\
	&& -{\frac{2\mathbb{M}^{2}}{\sqrt{\mathbb{M}^2-\mathcal{Q}}}{\rm arctanh} \left(1/2{\frac {2\mathbb{M}-2R_{{s}}}{\sqrt{\mathbb{M}^2-\mathcal{Q}}}}\right)}\nonumber\\
	&&+{\frac {\mathcal{Q}}{\sqrt{\mathbb{M}^2-\mathcal{Q}}}{\rm arctanh} \left(1/2{\frac {2\mathbb{M}-2R_{{s}}}{\sqrt{\mathbb{M}^2-\mathcal{Q}}}}\right)}.
\end{eqnarray}
\begin{figure}
\begin{center}
\includegraphics[width=0.38\textwidth]{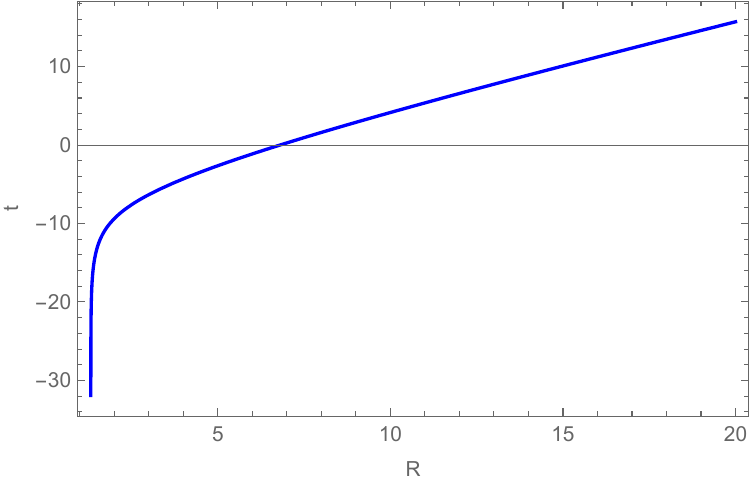}
\caption{Plot showing the dependence of $t$ on $R$}
\label{apparent_con2}
\end{center}
\end{figure} 
On putting $R=R_{\pm}$ in the above equation, the expression for time diverges. So, again, for an outside observer far away from the BH, the shell crossing takes an infinite amount of time. In this way, the non-singular BH interior with a radiation envelope evolves into a white hole, and future evolution depends on whether the white hole is stable or not. There is a single bounce if it is stable, and the white hole is the final state. If it is unstable, the expansion may halt, and again, a contraction process may be initiated, leading to BH formation with a non-singular interior, which may continue periodically. We shall investigate both possibilities in the near future.

As illustrated in Figure \ref{apparent_con2}, the post-bounce evolution is characterized by a smooth and monotonic increase of the coordinate time \( t \) with the radial coordinate \( R \). The absence of turning points or multivalued regions in the \( t(R) \) curve indicates that no shell-crossing singularities occur during the re-expansion phase. The matching of the interior collapsing region to the expanding branch across the bounce surface is achieved through continuous extrinsic curvature, ensuring a regular transition at the minimum radius. Consequently, the interior spacetime remains free from curvature singularities, and the collapse is replaced by a non-singular bounce leading to a regular expanding phase.

\section{Discussion} \label{conclusion}

In this paper, we have investigated the gravitational collapse of a dust fluid localized on a brane, within the framework of the SS braneworld scenario. In our model, the interior of the collapsing cloud is described by a Friedmann–Lemaître spacetime, representing a homogeneous and isotropic distribution of pressureless matter (dust). Surrounding this collapsing region, the exterior geometry is initially characterized by a Vaidya radiating envelope, which models the emission of null radiation during the collapse. As the collapse proceeds and the radiation phase ceases, the exterior geometry asymptotically approaches a static vacuum spacetime. This final exterior takes the form of a RN solution, modified by a positive tidal charge arising from the influence of the bulk geometry rather than from any physical electric charge.

A major challenge encountered in this analysis is the difficulty of performing a smooth matching between the interior Friedmann–Lemaître region and the final static RN exterior. This complication arises due to the breakdown of Birkhoff’s theorem in the braneworld context. Unlike in standard general relativity, where Birkhoff’s theorem ensures that the exterior of a spherically symmetric matter distribution must be static and uniquely given by the Schwarzschild (or RN) solution, the presence of extra dimensions in braneworld gravity leads to a richer structure of possible exterior solutions. In particular, bulk gravitational effects can introduce dynamical features in the exterior, preventing a straightforward or smooth junction with the interior collapsing region.

In the context of braneworld gravity, consider the collapse of a homogeneous dust fluid confined to the brane, where the standard matter energy-momentum tensor is that of pressureless dust. However, due to the embedding of the 4-dimensional brane in a higher-dimensional bulk spacetime, the effective gravitational field equations on the brane receive non-local corrections from the projection of the bulk Weyl tensor. These corrections manifest as an additional term, $W_{\mu\nu}$, in the effective Einstein equations, which behaves as a trace-free, radiation-like fluid with components including an effective energy density $U$, anisotropic stress $P_{\mu\nu}$, and energy flux $\mathbb{Q}_\mu$. Although the local brane matter is isotropic and pressureless, the projected bulk effects induce an inherent anisotropy in the effective stress-energy tensor through $P_{\mu\nu}$. This anisotropic stress, particularly under dynamical evolution such as collapse, can lead to the development of a non-zero radial energy flux component $\mathbb{Q}_\mu$, which is interpreted as an effective heat flux on the brane. As the collapsing dust evolves, this heat flux propagates outward, and upon reaching the boundary of the dust sphere, it can no longer be matched to a static Schwarzschild exterior. Instead, the appropriate exterior spacetime becomes the Vaidya solution, which describes an outgoing null radiation field. The radial heat flux at the boundary corresponds to the loss of mass from the system via radiation, with the rate of change of the Vaidya mass function determined by the magnitude of the flux. Thus, the inherent pressure anisotropy induced by non-local bulk gravitational effects can, through the generation of effective heat flux, lead to Vaidya-type radiation in a collapsing homogeneous dust configuration on the brane.

It is worth highlighting that in the Randall–Sundrum braneworld scenario, exact black hole solutions have been previously constructed on the brane~\cite{ND1}. In those cases, the RN metric emerges even in the absence of electromagnetic fields, with the role of electric charge effectively played by a ``tidal charge" induced from the bulk Weyl tensor. This tidal charge can take either sign, but a positive tidal charge, as considered in our work, leads to notable deviations from standard black hole spacetimes in four-dimensional general relativity.

In our analysis, we have demonstrated that as long as the braneworld effects remain significant, the effective brane tension remains finite. This, in turn, ensures that the four-dimensional scalar curvature on the brane remains regular throughout the collapse process. The finiteness of the scalar curvature indicates that no curvature singularity forms at the classical level, at least within the regime where braneworld corrections dominate. The interior BH dynamics should resemble that of a cyclic universe. The expansion factor will reach a minimum (given by Eq. \ref{e2})  depending on the brane tension and then bounce back as the collapse halts. The scale factor has been obtained numerically during the collapse in Fig. 2 as no exact solution is available for Eqns.  (\ref{e100})-(\ref{98}). However, in the post-collapse dynamics, the scale factor can be exactly described by Eqns. (\ref{BHEva})-(\ref{E24}) which again justifies the bounce. As the matter crossing the horizon will be trapped inside without ever reaching the singularity, the bounce can occur repetitively, producing cyclic behaviour in the interior of the BH.  This result offers an intriguing possibility for singularity avoidance due to higher-dimensional effects and suggests that braneworld gravity may provide a viable setting for exploring regular models of gravitational collapse.

In the scenario of gravitational collapse of a homogeneous dust sphere on a braneworld with a timelike extra dimension, a rich structure of non-local gravitational effects arises due to the embedding of the 4-dimensional brane in a higher-dimensional bulk. These bulk-induced corrections manifest on the brane as effective modifications to the matter content, captured by the effective energy density and anisotropic pressures. Although the strong energy condition is marginally satisfied by the effective matter, violations of the null, weak, and dominant energy conditions, particularly in the radial direction and on average, point to a significant departure from standard general relativistic behavior. Physically, such violations are deeply tied to the possibility of avoiding the formation of spacetime singularities. In this braneworld model, the gravitational collapse does not culminate in a singularity but instead halts or transitions smoothly, suggesting the emergence of a repulsive gravitational behavior at high curvatures. This feature aligns with similar phenomena observed in other quantum gravity-inspired models like loop quantum cosmology (LQC), where a repulsive quantum geometric force arises to counteract the infinite curvature that would otherwise develop.

Importantly, the presence of a timelike extra dimension in the braneworld plays a crucial role in this mechanism. Unlike a spacelike extra dimension, which typically contributes attractive corrections, a timelike extra dimension reverses the sign of certain geometric contributions from the bulk, effectively generating a repulsive force under extreme curvature conditions. This repulsive effect, emerging naturally from the geometrical structure of the theory, acts to counter gravitational collapse at high energy densities, thereby preventing the divergence of curvature invariants. The same framework that avoids singularity formation in gravitational collapse also allows for non-singular cosmological evolution, where the classical Big Bang singularity is replaced by a smooth bounce. Furthermore, this repulsive force contributes to the stability and traversability of wormholes in the braneworld, as it ensures that the tidal forces at the wormhole throat remain finite, thereby preventing divergence in the Weyl curvature and ensuring regular geometry throughout. Altogether, the timelike extra-dimensional braneworld scenario provides a compelling and consistent framework for addressing and potentially resolving classical curvature singularities across different physical contexts, collapse, cosmology, and wormholes, suggesting that the timelike nature of the extra dimension may encode a fundamental geometric mechanism for singularity avoidance in a manner analogous to but distinct from quantum gravity corrections.

\section*{Acknowledgments}
RS and CS express their sincere gratitude to Profs. P. Peter (IAP, Paris) and 
Varun Sahni (IUCAA, Pune)
for numerous suggestions and discussions instrumental 
in the development of this work.


\bibliography{Refs.bib}
\bibliographystyle{./utphys1}

\end{document}